\documentclass[preprint]{aastex63}
\usepackage{microtype}
\usepackage{natbib}
\usepackage{textcomp,gensymb}
\usepackage{xspace}
\usepackage{booktabs}

\newcommand{\sdo}{{\it SDO}\xspace}
\newcommand{\sta}{{\it STA}\xspace}
\newcommand{\stb}{{\it STB}\xspace}
\newcommand{\soho}{{\it SOHO}\xspace}
\newcommand{\goes}{{\it GOES}\xspace}
\newcommand{\hsi}{{\it RHESSI}\xspace}
\newcommand{\ie}{{\it i.e.}}
\newcommand{\eg}{{\it e.g.}}
\newcommand{\rs}{~$R_\odot$\xspace}
\newcommand{\kms}{~km~s$^{-1}$\xspace}
\newcommand{\kmss}{~km~s$^{-2}$\xspace}
\newcommand{\ai}{~\r{A}\xspace}

\received{May 22, 2020}
\revised{June 19, 2020}
\accepted{June 19, 2020}
\submitjournal{ApJL}

\shorttitle{Flare--CME Coupling}
\shortauthors{Gou et al.}

\begin{document}

\title{Solar Flare--CME Coupling Throughout Two Acceleration Phases of a Fast CME}

\correspondingauthor{Tingyu Gou}
\email{tygou@ustc.edu.cn}

\author[0000-0003-0510-3175]{Tingyu Gou}
\affiliation{CAS Key Laboratory of Geospace Environment, 
	Department of Geophysics and Planetary Sciences, \\
	University of Science and Technology of China, Hefei 230026, China}
\affiliation{CAS Center for Excellence in Comparative Planetology, \\
	University of Science and Technology of China, Hefei 230026, China}

\author[0000-0003-2073-002X]{Astrid M. Veronig}
\affiliation{Institute of Physics, University of Graz, A-8010 Graz, Austria} 
\affiliation{Kanzelh\"{o}he Observatory for Solar and Environmental
	Research, \\University of Graz, A-9521 Treffen, Austria}

\author[0000-0003-4618-4979]{Rui Liu}
\affiliation{CAS Key Laboratory of Geospace Environment, 
	Department of Geophysics and Planetary Sciences, \\
	University of Science and Technology of China, Hefei 230026, China}
\affiliation{CAS Center for Excellence in Comparative Planetology, \\
	University of Science and Technology of China, Hefei 230026, China}

\author[0000-0002-5996-0693]{Bin Zhuang}
\affiliation{Institute for the Study of Earth, Oceans, and Space, \\
	University of New Hampshire, Durham, NH 03824, USA} 

\author{Mateja Dumbovi\'c}
\affiliation{Institute of Physics, University of Graz, A-8010 Graz, Austria}
\affiliation{Hvar Observatory, Faculty of Geodesy, University of Zagreb, HR-10000 Zagreb, Croatia}

\author{Tatiana Podladchikova}
\affiliation{Skolkovo Institute of Science and Technology, Bolshoy Boulevard 30, bld. 1, Moscow 121205, Russia} 

\author{Hamish A. S. Reid}
\affiliation{SUPA School of Physics and Astronomy, University of Glasgow, Glasgow G12 8QQ, UK}
\affiliation{Department of Space and Climate Physics, University College London, London, RH5 6NT, UK}

\author{Manuela Temmer}
\affiliation{Institute of Physics, University of Graz, A-8010 Graz, Austria}

\author{Karin Dissauer}
\affiliation{Institute of Physics, University of Graz, A-8010 Graz, Austria}

\author{Bojan Vr\v{s}nak}
\affiliation{Hvar Observatory, Faculty of Geodesy, University of Zagreb, HR-10000 Zagreb, Croatia}

\author[0000-0002-8887-3919]{Yuming Wang}
\affiliation{CAS Key Laboratory of Geospace Environment, 
	Department of Geophysics and Planetary Sciences, \\ 
	University of Science and Technology of China, Hefei 230026, China}
\affiliation{CAS Center for Excellence in Comparative Planetology, \\
	University of Science and Technology of China, Hefei 230026, China}

\begin{abstract}
Solar flares and coronal mass ejections (CMEs) are closely coupled through magnetic reconnection. CMEs are usually accelerated impulsively within the low solar corona, synchronized with the impulsive flare energy release. We investigate the dynamic evolution of a fast CME and its associated X2.8 flare occurring on 2013 May 13. The CME experiences two distinct phases of enhanced acceleration, an impulsive one with a peak value of $\sim$5\kmss followed by an extended phase with accelerations up to 0.7\kmss. The two-phase CME dynamics is associated with a two-episode flare energy release.  While the first episode is consistent with the ``standard'' eruption of a magnetic flux rope, the second episode of flare energy release is initiated by the reconnection of a large-scale loop in the aftermath of the eruption and produces stronger nonthermal emission up to $\gamma$-rays. In addition, this long-duration flare reveals clear signs of ongoing magnetic reconnection during the decay phase, evidenced by extended HXR bursts with energies up to 100--300~keV and intermittent downflows of reconnected loops for $>$4~hours. The observations reveal that the two-step flare reconnection substantially contributes to the two-phase CME acceleration, and the impulsive CME acceleration precedes the most intense flare energy release. The implications of this non-standard flare/CME observation are discussed.
\end{abstract}


\section{Introduction}

Coronal mass ejections (CMEs) are clouds of magnetized plasma that erupt from the Sun's atmosphere and propagate into the interplanetary space. They are often accompanied by a large amount of magnetic energy release and can cause extreme space weather events when arriving at the Earth. Studies have revealed that CMEs usually undergo three stages of dynamic evolution: a slow rise, a fast acceleration, and a propagation phase \citep{ZhangJ2001,ZhangJ2004}. The final speed of a CME varies in a wide range of about 100--3500\kms \citep[\eg,][]{Gopalswamy2009,Lamy2019}, while its main acceleration usually takes place within a few to tens of minutes at low coronal heights \citep[\eg,][]{ZhangJ2001,Vrsnak2007,Temmer2008,Bein2011,Veronig2018}, where the Lorenz force that accounts for the liftoff of a CME is strong.

The launch of a CME is often accompanied by a rapid release of magnetic free energy in the solar atmosphere, in the form of flares that emit radiations across the entire electromagnetic spectrum. The two phenomena are closely coupled through magnetic reconnection. By reconfiguring magnetic filed lines, reconnection on one hand provides for the impulsive and vast energy release that is used for plasma heating and particle acceleration in flares \citep[see the review by][and references therein]{Shibata2011}, on the other hand facilitates the CME acceleration by reducing the tension of the overlying arcade and supplying additional poloidal magnetic flux to the erupting structure \citep[\eg,][]{Lin2000,Vrsnak2008}. The close relationship between flares and CMEs are presented in various studies by revealing a close temporal correlation between the flare soft X-ray (SXR) flux and the CME velocity profile \citep{ZhangJ2001,ZhangJ2004,Vrsnak2004}, and even the synchronization between the flare hard X-ray (HXR) emission and the CME acceleration \citep{Temmer2008,Temmer2010,BS2012}. A number of statistical studies \citep[\eg,][]{Maricic2007,Bein2011,BS2012,Cheng2020} provide further evidence for the close relation between the flare energy release and the CME impulsive acceleration, suggestive of a close link and feedback relationship between the CME dynamics and flare reconnection \citep{Temmer2010,Vrsnak2004,Vrsnak2008,Veronig2018}.

The positive feedback between the CME dynamics and the associated flare is established via magnetic reconnection in the current sheet (CS), which is most intense in the flare impulsive phase \citep[for observational signatures of the hot elongated CS, see, \eg,][]{Cheng2018,Warren2018,ChenB2020}. The vertical CS becomes observationally more prominent in the flare decay phase after the peak of the \goes SXR flux \citep[\eg,][]{LiuR2013}, which can usually be observed above the candle-flame-shaped flare loops that are considered as direct evidence of magnetic reconnection \citep[\eg,][]{Tsuneta1996,Lin2005,Gou2015,Gou2016}. Due to the magnetic tension force, these newly reconnected cusp-shaped field lines quickly retract to form more relaxed round-shape ones, which is widely known as the field line shrinkage \citep{Forbes1996,Vrsnak2006}. Meanwhile above the flare arcade, tadpole-like supra-arcade downflows (SADs) and bright supra-arcade downflow loops (SADLs) quickly descend from the reconnection site and merge into the dense flare loop region \citep[\eg][]{McKenzie1999,Savage2011,LiuW2013,Innes2014,ChenX2017}. Although the exact physical process is still unclear, it is believed that these features are closely related to the downward outflows of magnetic reconnection.

We study the dynamics of a fast CME and its relation to the associated X2.8 flare occurring on 2013 May 13. This event is the second X-class flare from NOAA active region 11748 on that day and has been studied before. \citet{MO2014} and \citet{SH2014} presented the unusual loop-prominence system in polarized white light that formed after the flare, indicative of high coronal densities. \citet{Gou2019} presented detailed observations of the buildup of the magnetic flux rope and large-scale CME from the coalescence of multiple small-scale plasmoids during the early stage of the flare. \citet{Gou2017} focused on the two distinct episodes of the flare energy release associated with two-step reconnection. The first episode is characterized by the ``standard'' flux rope eruption, and the second one is initiated by the reconnection of a loop leg behind the eruption, which leads to even stronger particle acceleration observed in emissions of high-energy HXRs and $\gamma$-rays.

In this letter, we concentrate on the dynamic evolution of the associated CME and find that the two strong episodes of the flare energy release are associated with two distinct phases of CME acceleration. We also find that the distribution of the CME acceleration and the flare nonthermal emission is different between the two phases, distinct from the standard flare model where they are supposed to be synchronized. To our knowledge, this is the first time that two distinct episodes of impulsive acceleration could be identified in a fast CME, suggestive of a different energy distribution from its associated flare.

\section{Data and Instruments}

We use the high spatial and temporal resolution EUV imagery (0.$''$6 and 12s) by the Atmospheric Imaging Assembly \citep[AIA;][]{Lemen2012} on board the Solar Dynamics Observatory \citep[\sdo;][]{Pesnell2012} to study the dynamic evolution of the eruption in the inner corona, and mainly focus on the 131\ai channel (primarily Fe XXI emission line, with a peak response temperature of log$T$=7.05). We use X-ray observations from the Reuven Ramaty High-Energy Solar Spectroscopic Imager \citep[\hsi;][]{LinRP2002} and the Gamma-ray Burst Monitor (GBM) onboard the Fermi Gamma-ray Space Telescope (Fermi hereafter), to study the energy release of the associated flare. The subsequent white-light CME is observed by the Solar and Heliospheric Observatory/Large Angle and Spectrometric Coronagraph \citep[\soho/LASCO, C2: 1.5--6\rs, C3: 3.7--30\rs;][]{Brueckner1995}, and the coronagraphs \citep[COR1: 1.5--4\rs and COR2: 2.5--15\rs;][]{Howard2008} on both the ``Ahead" and ``Behind'' satellites of the Solar Terrestrial Relations Observatory \citep[{\it STEREO};][\sta\ and \stb\ hereafter]{Kaiser2008}, which are about 136.3$\degree$ and 141.6$\degree$ separated from the Earth on 2013 May 13, respectively. In addition, radio observations obtained by \stb/WAVES \citep{Bougeret2008} are also included.

\section{Results}\label{sec:obs}

\subsection{Event Overview}

\begin{figure}[htbp]
	\centering
	\includegraphics[width=\textwidth]{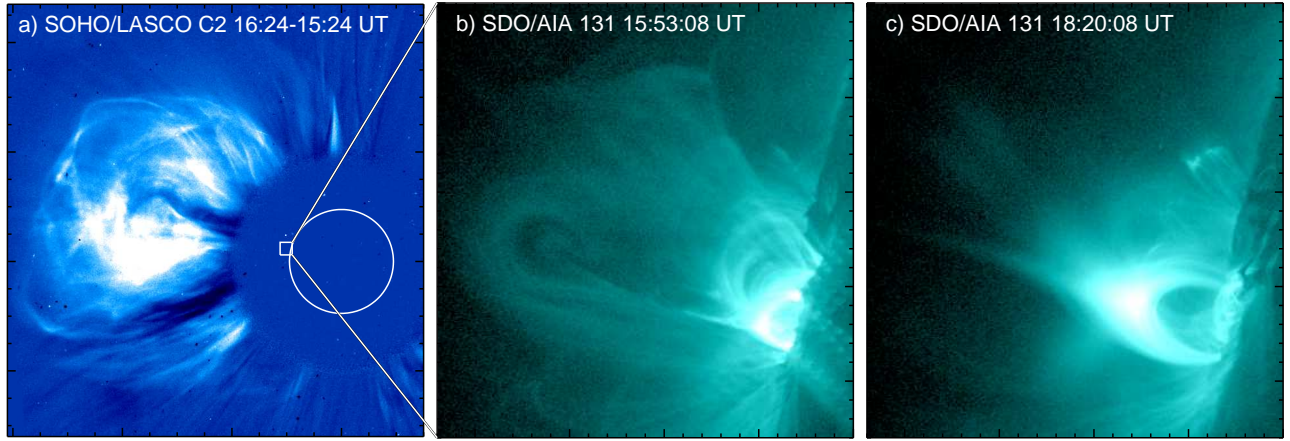}
	\caption{\small Solar eruption on 2013 May 13 observed by \soho/LASCO and \sdo/AIA 131\ai. In panel (a), the white circle indicates the solar limb, and the rectangular shows the field of view (FOV) of panels (b,c).} 
	\label{fig:ov}
\end{figure}

The event under study originates from NOAA active region 11748 near the northeast solar limb on 2013 May 13 (Figure~\ref{fig:ov}). It manifests as the eruption of a magnetic flux rope as observed in the \sdo/AIA 131\ai filter (see Figure~\ref{fig:ov}, also \citealp{Gou2019} for details), the bottom of which connects the cusp-shaped flare loops underneath by the vertical CS, in good accordance with the standard model of solar eruptions (\eg, \citealp{Carmichael1964,Sturrock1966,Hirayama1974,KP1976}; see also reviews by \citealp{Shibata2011,Holman2012}). The eruption produces a fast halo CME with a velocity of $\sim$1850\kms (according to the \soho/LASCO CME catalog, \url{http://cdaw.gsfc.nasa.gov/CME_list/}), and an intense long-duration X2.8 flare that starts at 15:48~UT and peaks at 16:05~UT. This flare is associated with strong particle acceleration as observed in emissions of high-energy HXRs and $\gamma$-rays (see details in \citealp{Gou2017}). Here we concentrate on the dynamic evolution of the CME in the solar corona, especially on the impulsive acceleration process and its relation to the flare energy release and high-energy particle acceleration.

\subsection{CME Dynamics}

\begin{figure}[htbp]
	\centering
	\includegraphics[width=\textwidth]{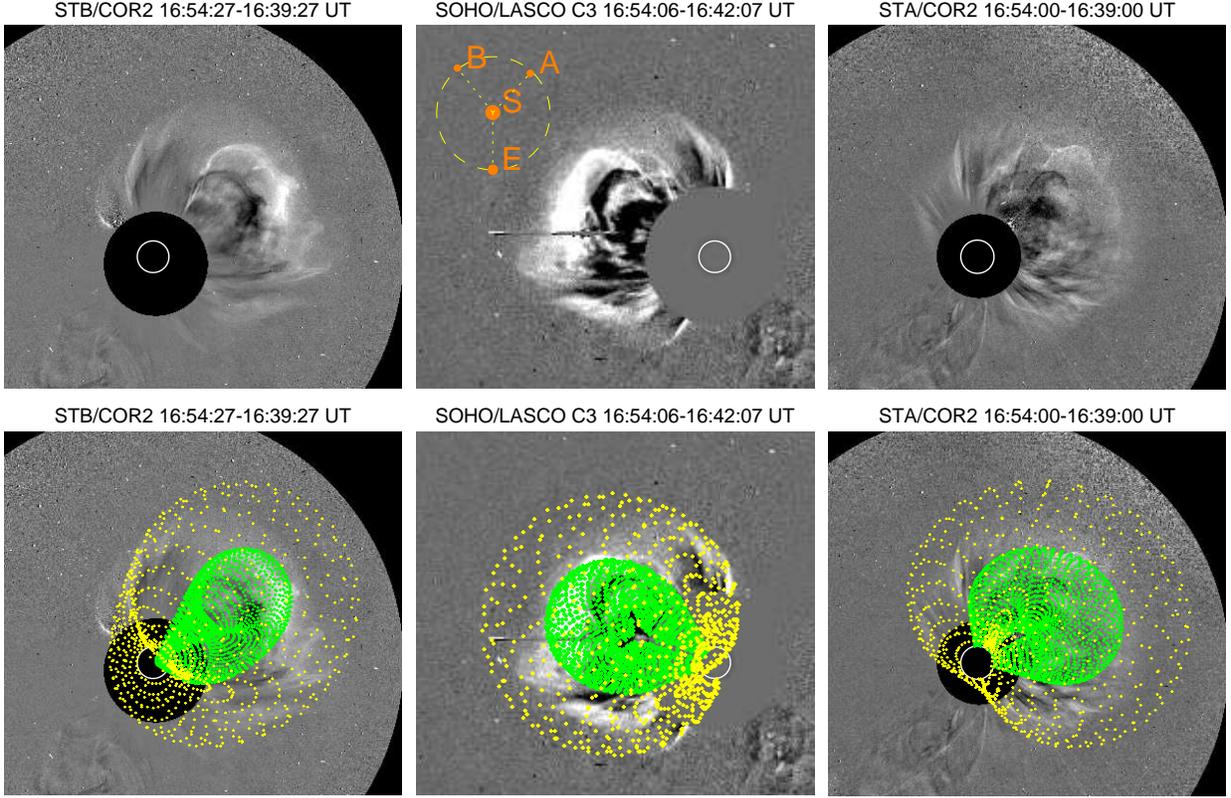}
	\caption{\small GCS reconstruction of the CME (green) and the shock (yellow) at $\sim$16:54 UT using simultaneous observations by \stb/COR2 (left), LASCO/C3 (middle), and \sta/COR2 (right). The inset in the top middle panel shows the positions of \sta (A) and \stb (B) relative to the Sun (S) and the Earth (E) on 2013 May 13.}
	\label{fig:gcs}
\end{figure}

\begin{figure}[htb]
	\centering
	\includegraphics[width=\textwidth]{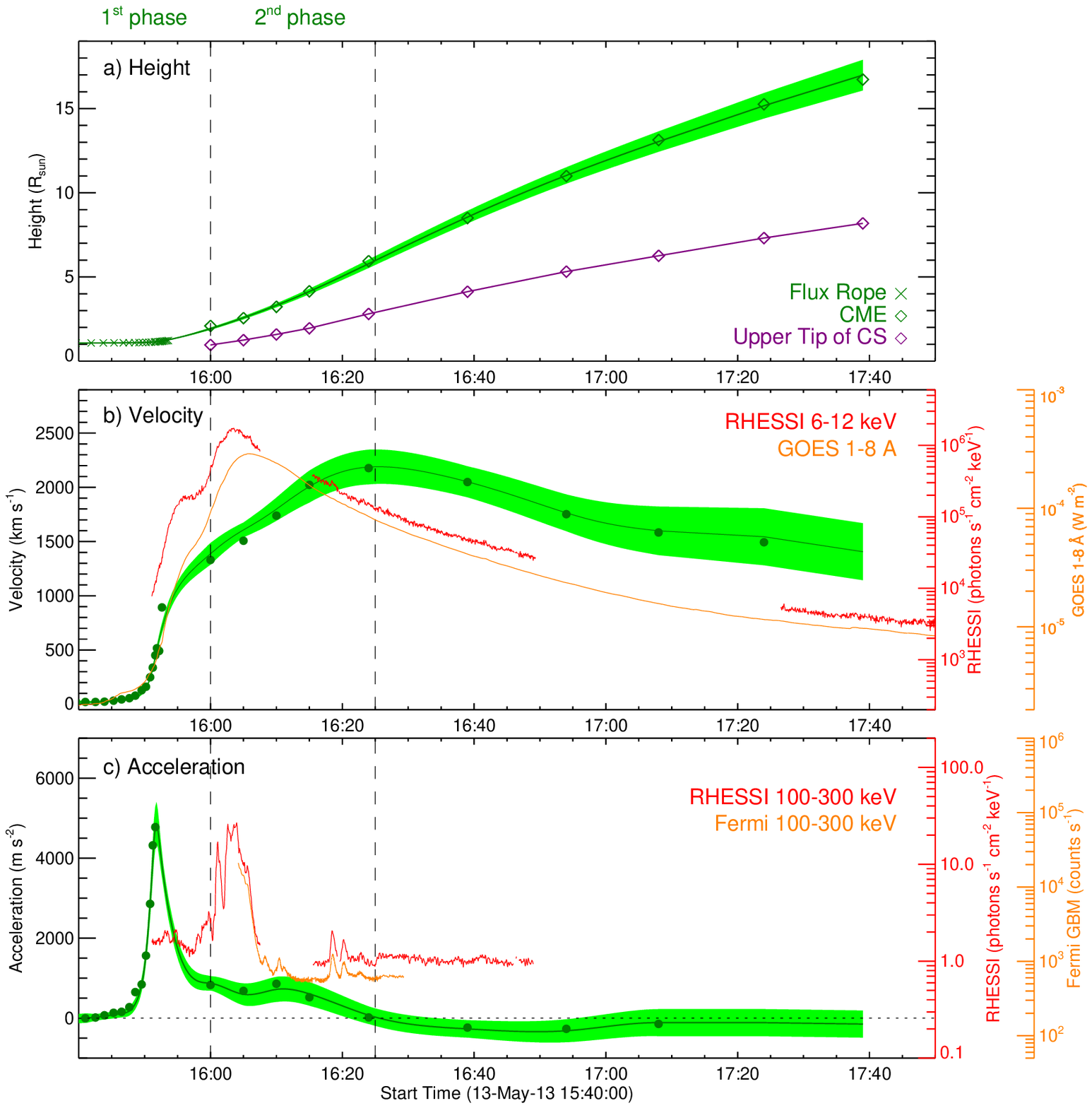}
	\caption{\small CME kinematics (scaled by the left y-axes) compared with flare X-ray fluxes (scaled by the right y-axes). The green cross symbols in panel (a) indicate the height of the flux rope front observed in AIA 131\ai images. The colored diamonds in panel (a) indicate the heights of the CME apex (dark green) and the lower boundary of CME (purple) measured in the GCS model. All heights are with respect to the solar center. The dark green dots in panels (b,c) indicate direct numerical derivatives of the measured data points in panel (a). The dark green curves in (a--c) are smoothed kinematic profiles, with errors indicated by light green shadows. \goes, \hsi and Fermi X-ray fluxes are plotted in panels (b,c) in different colors as indicated by the legends. Fermi GBM observations from 16:04--16:30~UT are added to fill in the \hsi data gap between 16:07--16:16~UT when it crosses the South Atlantic Anomaly (SAA). The two vertical dashed lines mark the two acceleration phases of the CME. }
	\label{fig:kine}
\end{figure}
 
\sdo/AIA observes the eruption in the inner corona at the northeast limb, which allows us to study its dynamics with least projection effects. We measure the leading front of the magnetic ejecta in AIA 131\ai images to obtain its height-time evolution. The white-light CME in the outer corona is observed after 15:55~UT by the coronagraphs on board \sta, \stb\ and \soho from three different viewpoints (Figure \ref{fig:gcs}). This fast CME drives a shock in front of it. Based on the stereoscopic observations, we use the Graduated cylindrical shell \citep[GCS;][]{Thernisien2006,Thernisien2009} model to reconstruct the three-dimensional morphology of the eruption (Figure \ref{fig:gcs}). The model assumes a flux rope structure of the CME, and it is determined by three geometry and three position parameters: the aspect ratio $\kappa$, the half-angle $\alpha$ and the tilt of the croissant representing the CME, the longitude and the latitude of the source region, and the height of the CME apex. We also model the shock in front of the CME by geometrically reproducing a sphere ($\alpha$ = 0, $\kappa$ = 1). Figure \ref{fig:kine}(a) shows the obtained height-time evolution of the CME (measured along the apex of the GCS fitting), as well as the lower boundary of the flux rope in the model, which basically corresponds to the upper tip of the CS underneath the erupting CME. 

We combine the height-time measurements to study the complete kinematics of the CME in the corona (Figure \ref{fig:kine}). To derive the velocity and acceleration profiles, we first smooth the height-time data and derive the first and second time derivatives. The smoothing algorithm is based on the method described in \citet{Podl2017} and applied in \citet{Dissauer2019}, extended toward non-equidistant data. From the obtained acceleration profiles, we further interpolate to equidistant data points based on the minimization of the second derivatives and reconstruct the corresponding velocity and height profiles by integration. We also obtain the errors of kinematic profiles by assuming that the measurement errors of heights amount to 6 AIA pixels in the inner corona and 3\% of GCS heights in the outer corona. Figure \ref{fig:kine}(b,c) shows the obtained velocity and acceleration profiles (dark green curves with errors in light green), and they are well aligned with the direct numerical derivatives of the data points (dark green dots). The magnetic ejecta starts to accelerate at $\sim$15:41~UT, several minutes earlier than the \goes flare start (15:48~UT), suggestive of the role of an ideal instability to trigger the eruption. The CME achieves its highest velocity of 2190($\pm$158)\kms at 16:25~UT (\ie, 20~min after the peak of the \goes SXR flux) at a height of 6.07($\pm$0.26)\rs (with respect to the solar center). For comparison, the CME velocity at the flare peak time (\ie, 16:05~UT) is 1607($\pm$68)\kms at a height of 2.58($\pm$0.09)\rs. 

The acceleration of the CME exhibits two distinct phases, an impulsive peak (15:41--16:00~UT) followed by an enhanced gradual phase (16:00--16:25~UT), as marked by the vertical dashed lines in Figure \ref{fig:kine}. The first phase of impulsive acceleration achieves a peak value of 4.88($\pm$0.52)\kmss at 15:52~UT, when the CME is at a height of 1.15($\pm$0.01)\rs. The second phase undergoes an extended acceleration of several hundreds m s$^{-2}$ (up to 0.73($\pm$0.31)\kmss at 16:11~UT), and it raises the CME velocity from 1387($\pm$100)\kms at a height of 1.93($\pm$0.07)\rs to 2190($\pm$158)\kms at 6.07($\pm$0.26)\rs, which is considerable as compared to the velocity increment during the first phase. We study the mechanism of the CME acceleration and give interpretations in terms of the flare reconnection in the following sections.

\subsection{Flare Energy Release}

\subsubsection{Two-Step Reconnection}

\begin{figure}[htbp]
	\centering
	\includegraphics[width=0.95\textwidth]{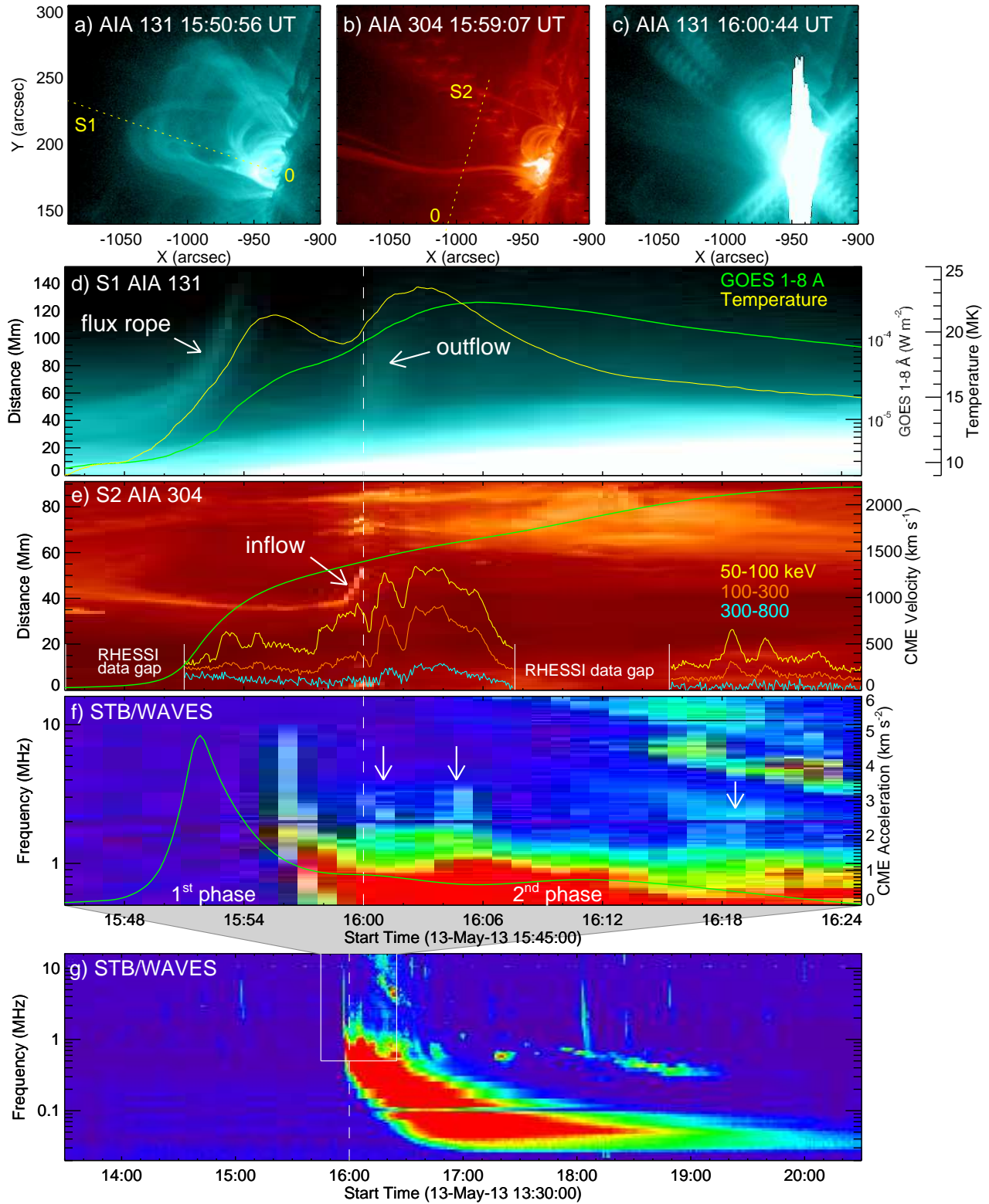}
	\caption{\small Two acceleration phases of the CME. a--c) \sdo/AIA 131 and 304\ai images showing the flux rope eruption, loop-leg inflow and upward outflow of the second-step reconnection. d,e) Stack plots derived from \sdo/AIA 131 and 304\ai imagery along slits S1 and S2 as shown in panels (a,b) (the starts are labeled as ``0''). f,g) Dynamic radio spectra observed by \stb/WAVES. The yellow and light green curves in panels (d--f) (scaled by the right y-axes) represent the \goes plasma temperature, \goes 1--8\ai flux, CME velocity and acceleration profiles, respectively. \hsi photon fluxes are plotted in panel (e) on arbitrary y-axes. The white arrows in panel (f) mark several injections of type III radio bursts. The vertical dashed line in panels (d--g) marks the onset of the second-step reconnection. }
	\label{fig:rec}
\end{figure}

The \goes X2.8 flare associated with the CME experiences two distinct episodes of energy release attributed to a two-step magnetic reconnection process, as reported by \citet{Gou2017}. Here we summarize some observational evidences in Figure \ref{fig:rec}(a--e) so as to compare with the CME dynamics. One can see that after the first step of flux rope eruption that exhibits typical characteristics of an eruptive flare--CME event, the second-step reconnection is directly imaged by \sdo/AIA at $\sim$16:00~UT, manifesting as the disappearance of the cool loop-leg inflow in 304\ai and simultaneous fast outflows of hot plasma indicative of reconnection outflow jets. The second episode of energy release is associated with even stronger particle acceleration than the first one, evidenced by stronger bursts of HXR emissions and even $\gamma$-rays (Figure~\ref{fig:rec}(e)). In addition, \stb/WAVES observes significant injections and increases of type III radio emission at $\sim$3~MHz (Figure \ref{fig:rec}(f,g)) that coincide with the \hsi HXR bursts after 16:00~UT, indicative of an increased number of accelerated electrons escaping from the Sun. The timing confirms that the second-step reconnection is associated with strong acceleration of electrons, which propagate both downward to the chromosphere to emit in HXRs by the nonthermal bremsstrahlung mechanism and upward into the interplanetary space to excite fast-drift type III radio burst. \stb/WAVES also observes decameter-hectometric type II radio bursts (Figure~\ref{fig:rec}(f,g); fundamental and harmonic components at frequencies between 16~MHz and 3~MHz from 16:10 to 16:30~UT, and later one component at about 0.9-0.3~MHz from 17:50 to 19:20~UT), which provide evidences for the propagation of the shock driven by the fast CME.

We note that the two strong episodes of flare energy release are temporally related to the two phases of CME acceleration. The resultant two-phase evolution of the CME velocity are also associated with a two-episode enhancement of the \goes SXR flux as well as that of flare temperature evolution (Figure~\ref{fig:rec}(d--e)). It implies that the second-step reconnection not only gives rise to another stronger episode of energy release in the flare, but also contributes to an additional phase of the CME acceleration, impulsively increasing its speed beyond the main phase.

\subsubsection{Ongoing Reconnection in Flare Decay Phase}

\begin{figure}[htbp]
	\centering
	\includegraphics[width=\textwidth]{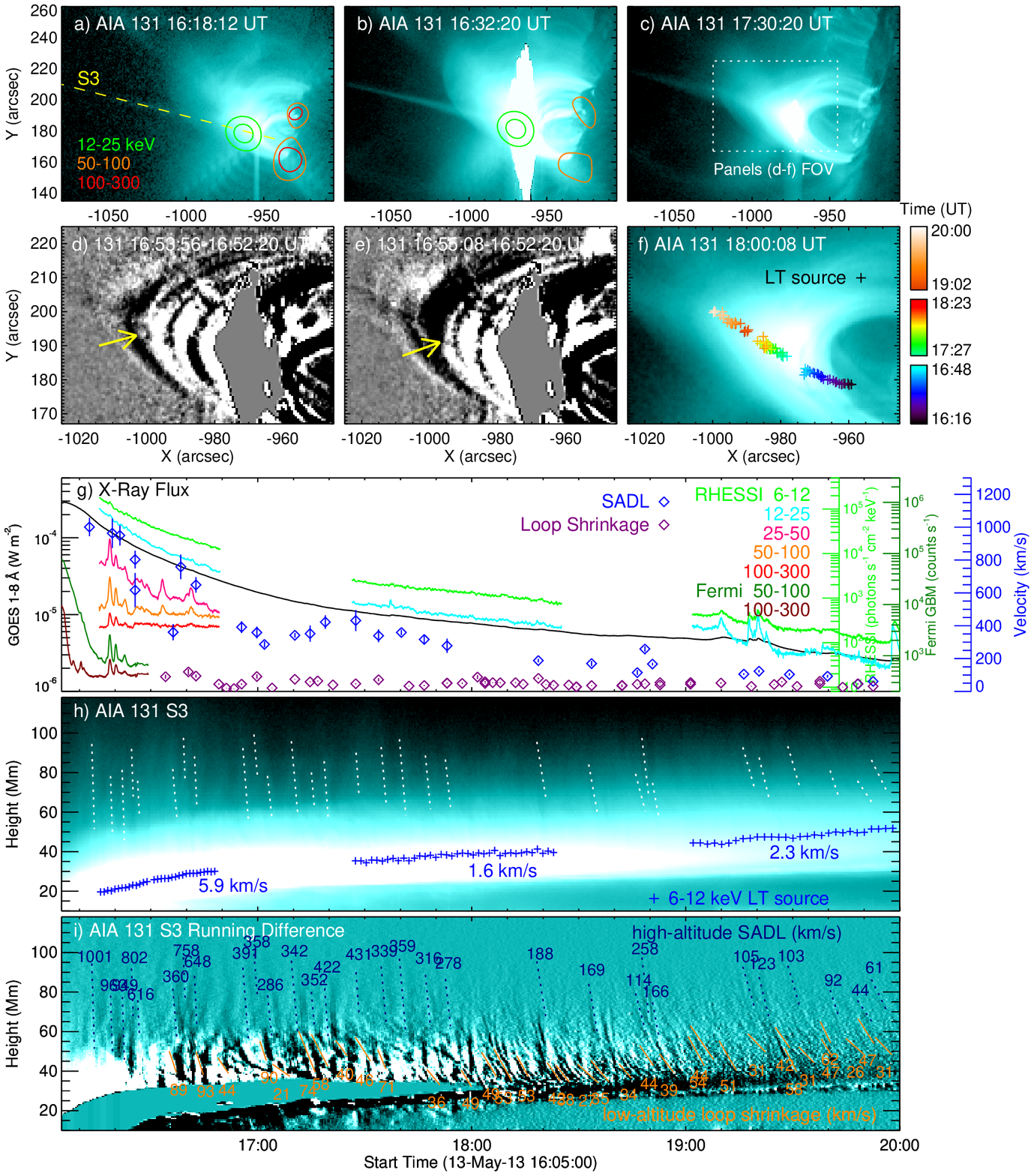}
	\caption{\small Flare dynamics in the decay phase. a--f) Snapshots of \sdo/AIA 131 \r{A} images. \hsi\ HXR sources observed at $\sim$16:18~UT and $\sim$16:32~UT are overlaid in panels (a,b). The plus signs in panel (f) indicate the location of 6--12~keV LT sources in three \hsi observation windows during the flare decay phase. g) \goes, \hsi, and Fermi GBM X-ray fluxes. The curves are vertically shifted to avoid overlap. h,i) Stack plots of slit S3 in panel (a), the y-axes indicate the height above the solar limb. The blue plus symbols in panel (h) indicate the heights of the \hsi LT source, labeled with rising speeds of linear fits for each time period. The dotted lines in panels (h,i) indicate the linear fits of various tracks left by the moving loops, labeled with the resultant speed in \kms. The speeds for individual retracting loops are also plotted in panel (g) with blue and purple diamonds, scaled by the right y-axis.
	An animation is available for the \sdo/AIA 131\ai observations and its running difference images from 16:00~UT to 21:00~UT on 2013 May 13.}
	\label{fig:decay}
\end{figure}

After peaking at 16:05~UT, the flare experiences a long decay phase that lasts over 4 hours (see Figure \ref{fig:decay} and its animation). During this stage, the vertical CS and the cusp-shaped flare loops underneath become prominent, and the whole post-flare loop system grows higher as observed in \sdo/AIA 131\ai images. The growth is also evidenced by the temporal evolution of the \hsi loop-top (LT) source location in Figure~\ref{fig:decay}(f,h), which shows that the post-flare loop system grows higher with a speed of several\kms throughout the flare decay phase. Different from the apparent rising of the loop system, a multitude of individual post-flare loops are observed to contract downward toward the solar surface (see Figure \ref{fig:decay}(d,e) and its animation), indicative of the shrinkage of newly reconnected field lines \citep{Forbes1996,Priest2002,Vrsnak2006}. In the upper CS region, the SADLs are observed to move downward rapidly and merge into the dense flare LT underneath.

We place a virtual slit across the flare LT along the CS (Figure~\ref{fig:decay}(a); with a width of 8 AIA pixels) to study in detail the dynamics during the flare decay phase. In the generated stack plots in Figure~\ref{fig:decay}~(h,i), one can see the downward motion of the high-altitude SADLs and the shrinkage of low-altitude flare loops, both of which can be clearly identified until 20:00~UT. For the former, the speed in the early decay phase is about 1000\kms, higher than many earlier reportings \citep[\eg,][]{Savage2011,Innes2014} and comparable to the typical Alfv\'en speed in the active region corona that determines the local reconnection outflow speed. After 19:00~UT, the speed decreases to $\sim$100\kms, and its temporal evolution (see the blue diamonds in Figure~\ref{fig:decay}(g)) is similar to that of the Alfv\'en speed distribution above solar active regions, which starts to decrease at a height of $\sim$4\rs \citep[\eg, see Figure~6 in][]{Mann2003}.  For the latter, the speeds are about 20--95\kms, much smaller than those of the SADLs and consistent with the earlier measurements \citep[\eg,][]{LiuW2013}. One can see that these two kinds of motions almost merge together with similar speeds in the late decay phase (Figure~\ref{fig:decay}~(h,i)) as the reconnection site rises high enough. This finding supports the idea that they may correspond to different stages of the contraction of newly reconnected loops, which always shrink fastest at the moment they are formed and released from the reconnection region, and thereafter decelerate when approaching the rising flare loop system \citep[see also][]{LinJ2004,LiuW2013}. All these characteristic features identified, \ie, the dynamic CS, the growth of the post-flare loop system, fast retraction of SADLs, and the intermittent shrinkage of post-flare loops, provide clear indications that there is still very efficient magnetic reconnection in process during the extended decay phase of the flare. 

Another strong evidence for ongoing reconnection is the significant energy release in the form of strong HXR bursts that are recorded by both \hsi and Fermi GBM (Figure~\ref{fig:decay}(g)). One can see that as the SXR flux decreases after the flare peak, Fermi GBM records two groups of HXR bursts at 16:09--16:11~UT and 16:18--16:22~UT, at energies up to 100-300~keV. Similarly, \hsi records a group of several HXR bursts at 16:18--16:22~UT up to 100--300~keV. In addition, \hsi shows HXR bursts occurring until 16:45~UT up to energies of 50--100~keV. By reconstructing X-ray images, we find that the high-energy HXR emissions (50-300~keV) mainly originate from two footpoints of flare loops (Figure \ref{fig:decay}(a,b)), indicative of accelerated electrons moving downward to the chromosphere. Nearly at the same time of the HXR bursts detected at around 16:18~UT, one can see a large increase in the type III radio emission at around 3~MHz (Figure~\ref{fig:rec}(e--g)). It suggests an increased number of accelerated electrons escaping from the Sun along open field lines, which finally merge into a large branch of an interplanetary type III burst (Figure~\ref{fig:rec}(g)). The HXR and radio emissions further indicate that there is still efficient particle acceleration even beyond the flare rise phase, most probably caused by the ongoing magnetic reconnection process.

This can nicely explain the second CME acceleration period during the flare decay phase, which is well in line with the extended HXR bursts (Figure \ref{fig:kine}). Namely, the ongoing magnetic reconnection beyond the flare rise phase is still coupled to the impulsive acceleration of the associated fast CME up to a height of $\sim$6 $R_\odot$.

\section{Discussion and Conclusion}\label{sec:dis}

We investigate the dynamic evolution of a fast CME, which experiences two distinct phases of enhanced acceleration, \ie, an impulsive phase with a peak value of around 5\kmss and an additional gradual phase with extended acceleration up to $\sim$0.7\kmss. The associated X2.8 flare exhibits two strong episodes of energy release associated with two-step reconnection, which coincide with the two phases of the CME dynamic evolution. Notably, the second phase of flare energy release and high-energy particle acceleration is substantially stronger than the first one and shows nonthermal emissions even in the $\gamma$-ray range. In addition, this long-duration flare reveals clear signs of ongoing magnetic reconnection during its long decay phase, evidenced by efficient particle acceleration in the form of high-energy (up to 100--300~keV) HXR emission, and by prolonged ($>$4~hours) downflows of reconnected loops (SADLs),  shrinkage of post-flare loops, and continuous growth of the post-flare loop system observed by both \hsi and \sdo/AIA.

We note that the CME is accelerated as fast as 5\kmss during the first phase. This acceleration is among the highest values of CME accelerations ever reported \citep[\eg, see statistical results in][]{ZhangJ2006,Vrsnak2007,Bein2011}. Considering that the intense CME acceleration is facilitated by the flare reconnection that converts the magnetic flux confining the eruptive flux rope into the rope's own flux  \citep{Temmer2010,Vrsnak2008,Veronig2018}, we suggest that a significant part of the magnetic ejecta in this event is formed during the eruption by reconnection of the overlying magnetic arcade. This generally agrees with observations in \citet{Gou2019} that the large-scale CME builds up from a small-scale seed during its impulsive rising. Also, we note that this is consistent with the recent findings for the X8.2 flare/CME on 2017 September 10, where it was clearly shown that the CME core observed in the white-light coronagraph is due to frozen-in plasma added to the rising flux rope by magnetic reconnection in the current sheet below and not due to the erupting prominence material \citep{Veronig2018}.

Moreover, the second phase of the CME acceleration in this event is substantial, and to our knowledge this is the first case that features two distinct phases of significant CME acceleration. The peak value during the second phase is $>$0.7\kmss, which still falls into the top $\sim$30\% of the main peak accelerations of impulsive flare-associated CMEs in statistical studies \citep[\eg,][]{Bein2011}. Such an extended acceleration phase contributes to $\sim$36\% of the CME velocity, even though it is substantially weaker than the first impulsive peak, because the coronal magnetic field magnitude decreases rapidly as the CME rises into the outer corona. Considering that the full-fledged CME carries much more mass than during the first phase, the change rate of the momentum may be comparable. Thus, this second-phase acceleration with high values and a comparable duration to the first phase is distinct from the residual CME acceleration following the main phase \citep{ZhangJ2006,Cheng2010}, which generally exhibits much lower values (these may be positive or negative, with maximum values only up to several tens of m~s$^{-2}$). As a result of the two-phase acceleration, the CME finally reaches its peak velocity at a height of $>$6\rs from the solar center. This is among the upper range of coronal heights where the main acceleration of impulsive flare-associated CMEs typically ends \citep[\eg,][]{Bein2011}.

We observe that the second-step magnetic reconnection with rapid curve-in of the loop leg not only initiates a stronger episode of the flare energy release than the first one, but also contributes to a second phase of the CME acceleration. Magnetic reconnection beneath the CME enhances its acceleration by reducing the downward tension of the overlying field and at the same time increasing the upward magnetic pressure gradient by supplying additional poloidal magnetic flux into the CME \citep[\eg,][]{LinJ2004,Vrsnak2008}. We can see a time lag between the second episode of flare energy release shown as high-energy HXR emission (peaking at 16:04~UT) and the second phase of the CME acceleration (centered around 16:10~UT; Figure~\ref{fig:kine}(c)). Considering the role of magnetic reconnection in the CME acceleration by feeding magnetic flux, this would generally correspond to the time that the reconnection outflow jets need to reach the lower part of the erupting flux rope on Alfv\'enic time scales. According to the observation, the reconnection site at the time of the second-step reconnection is located low in the corona, about 20~Mm above the solar limb (see Figure~\ref{fig:rec}, where the loop-leg inflow is swept in and outflow plasmas originate; also details in \citealp{Gou2017}), and the height of the flux rope's lower boundary is measured as $\sim$0.6\rs by the GCS model (Figure~\ref{fig:kine}(a)). If we assume that the distance between these two is 0.5\rs, and the Alfv\'en speed is in the order of 1000\kms (as inferred from the speeds of SADLs in Figure~\ref{fig:decay}), the time delay is about 6~minutes. This is generally consistent with the observation. Thus, for the second acceleration phase when the CME runs far out and keeps moving fast, the accelerating effect of feeding new flux to the CME will be reduced and delayed. This can also explain why we observe that the second phase of the CME acceleration is generally an extended phase of enhanced acceleration, other than a sharp peak during the first phase when the flux rope is still located low in the corona. 

In particular, one can see a different distribution of the CME acceleration and the flare nonthermal emission during the two phases (\eg, Figure~\ref{fig:kine}). While the impulsive CME acceleration occurs in the first phase, the flare is associated with much more efficient acceleration of high-energy particles in the second phase. We note that in the second phase, although the CME runs far out, the reconnection still occurs at low altitudes, which is capable to accelerate large numbers of electrons needed for intense HXR emission. On the other hand, the CME acceleration during the second phase is weaker than what would be expected for such strong HXR emission, which could be attributed either to the weaker Lorenz force at larger coronal heights or to the larger CME inertia that increases with time. The observation shows a different scenario from the synchronization between the flare HXR emission and the CME acceleration as supposed in the standard model, and suggests a different energy distribution between the flare and the CME in the two phases.

In conclusion, the two strong episodes of energy release in this flare are associated with two distinct phases of the CME acceleration, and the impulsive CME acceleration occurs at an earlier stage than the peak of flare nonthermal emission. This unusual two-phase evolution finally produces a very fast CME and an intense long-duration X-class flare with $\gamma$-ray emission, and is suggestive of a coupling between the flare energy release and the CME acceleration during the two phases but with different energy distributions among the two phenomena.

\acknowledgments
We thank the \sdo, \hsi, \soho and \textit{STEREO} teams for the data. T.G., R.L., and Y.W. acknowledge the support by NSFC (Grant No. 11903032, 41761134088, 41774150, and 11925302), CAS Key Research Program (Grant No. KZZD-EW-01-4), the fundamental research funds for the central universities, and the Strategic Priority Program of the Chinese Academy of Sciences (Grant No. XDB41000000). A.M.V. acknowledges the support by the Austrian Science Fund (FWF): P27292-N20. B.V. and M.D. acknowledge funding from the EU H2020 grant agreement No. 824135 (SOLARNET) and support by the Croatian Science Foundation under the project 7549 (MSOC). H.A.S.R. was supported by the STFC grant ST/P000533/1.


\begin{thebibliography}{}
	\expandafter\ifx\csname natexlab\endcsname\relax\def\natexlab#1{#1}\fi
	\providecommand{\url}[1]{\href{#1}{#1}}
	\providecommand{\dodoi}[1]{doi:~\href{http://doi.org/#1}{\nolinkurl{#1}}}
	\providecommand{\doeprint}[1]{\href{http://ascl.net/#1}{\nolinkurl{http://ascl.net/#1}}}
	\providecommand{\doarXiv}[1]{\href{https://arxiv.org/abs/#1}{\nolinkurl{https://arxiv.org/abs/#1}}}
	
	\bibitem[{{Bein} {et~al.}(2011){Bein}, {Berkebile-Stoiser}, {Veronig},
		{Temmer}, {Muhr}, {Kienreich}, {Utz}, \& {Vr{\v{s}}nak}}]{Bein2011}
	{Bein}, B.~M., {Berkebile-Stoiser}, S., {Veronig}, A.~M., {et~al.} 2011, \apj,
	738, 191, \dodoi{10.1088/0004-637X/738/2/191}
	
	\bibitem[{{Berkebile-Stoiser} {et~al.}(2012){Berkebile-Stoiser}, {Veronig},
		{Bein}, \& {Temmer}}]{BS2012}
	{Berkebile-Stoiser}, S., {Veronig}, A.~M., {Bein}, B.~M., \& {Temmer}, M. 2012,
	\apj, 753, 88, \dodoi{10.1088/0004-637X/753/1/88}
	
	\bibitem[{{Bougeret} {et~al.}(2008){Bougeret}, {Goetz}, {Kaiser}, {Bale},
		{Kellogg}, {Maksimovic}, {Monge}, {Monson}, {Astier}, {Davy}, {Dekkali},
		{Hinze}, {Manning}, {Aguilar-Rodriguez}, {Bonnin}, {Briand}, {Cairns},
		{Cattell}, {Cecconi}, {Eastwood}, {Ergun}, {Fainberg}, {Hoang}, {Huttunen},
		{Krucker}, {Lecacheux}, {MacDowall}, {Macher}, {Mangeney}, {Meetre},
		{Moussas}, {Nguyen}, {Oswald}, {Pulupa}, {Reiner}, {Robinson}, {Rucker},
		{Salem}, {Santolik}, {Silvis}, {Ullrich}, {Zarka}, \&
		{Zouganelis}}]{Bougeret2008}
	{Bougeret}, J.~L., {Goetz}, K., {Kaiser}, M.~L., {et~al.} 2008, \ssr, 136, 487,
	\dodoi{10.1007/s11214-007-9298-8}
	
	\bibitem[{{Brueckner} {et~al.}(1995){Brueckner}, {Howard}, {Koomen},
		{Korendyke}, {Michels}, {Moses}, {Socker}, {Dere}, {Lamy}, {Llebaria},
		{Bout}, {Schwenn}, {Simnett}, {Bedford}, \& {Eyles}}]{Brueckner1995}
	{Brueckner}, G.~E., {Howard}, R.~A., {Koomen}, M.~J., {et~al.} 1995, \solphys,
	162, 357, \dodoi{10.1007/BF00733434}
	
	\bibitem[{{Carmichael}(1964)}]{Carmichael1964}
	{Carmichael}, H. 1964, {A Process for Flares}, Vol.~50, 451
	
	\bibitem[{{Chen} {et~al.}(2020){Chen}, {Shen}, {Gary}, {Reeves}, {Fleishman},
		{Yu}, {Guo}, {Krucker}, {Lin}, {Nita}, \& {Kong}}]{ChenB2020}
	{Chen}, B., {Shen}, C., {Gary}, D.~E., {et~al.} 2020, arXiv e-prints,
	arXiv:2005.12757.
	\newblock \doarXiv{2005.12757}
	
	\bibitem[{{Chen} {et~al.}(2017){Chen}, {Liu}, {Deng}, \& {Wang}}]{ChenX2017}
	{Chen}, X., {Liu}, R., {Deng}, N., \& {Wang}, H. 2017, \aap, 606, A84,
	\dodoi{10.1051/0004-6361/201629893}
	
	\bibitem[{{Cheng} {et~al.}(2018){Cheng}, {Li}, {Wan}, {Ding}, {Chen}, {Zhang},
		\& {Liu}}]{Cheng2018}
	{Cheng}, X., {Li}, Y., {Wan}, L.~F., {et~al.} 2018, \apj, 866, 64,
	\dodoi{10.3847/1538-4357/aadd16}
	
	\bibitem[{{Cheng} {et~al.}(2010){Cheng}, {Zhang}, {Ding}, \&
		{Poomvises}}]{Cheng2010}
	{Cheng}, X., {Zhang}, J., {Ding}, M.~D., \& {Poomvises}, W. 2010, \apj, 712,
	752, \dodoi{10.1088/0004-637X/712/1/752}
	
	\bibitem[{{Cheng} {et~al.}(2020){Cheng}, {Zhang}, {Kliem}, {T{\"o}r{\"o}k},
		{Xing}, {Zhou}, {Inhester}, \& {Ding}}]{Cheng2020}
	{Cheng}, X., {Zhang}, J., {Kliem}, B., {et~al.} 2020, \apj, 894, 85,
	\dodoi{10.3847/1538-4357/ab886a}
	
	\bibitem[{{Dissauer} {et~al.}(2019){Dissauer}, {Veronig}, {Temmer}, \&
		{Podladchikova}}]{Dissauer2019}
	{Dissauer}, K., {Veronig}, A.~M., {Temmer}, M., \& {Podladchikova}, T. 2019,
	\apj, 874, 123, \dodoi{10.3847/1538-4357/ab0962}
	
	\bibitem[{{Forbes} \& {Acton}(1996)}]{Forbes1996}
	{Forbes}, T.~G., \& {Acton}, L.~W. 1996, \apj, 459, 330, \dodoi{10.1086/176896}
	
	\bibitem[{{Gopalswamy} {et~al.}(2009){Gopalswamy}, {Yashiro}, {Michalek},
		{Stenborg}, {Vourlidas}, {Freeland}, \& {Howard}}]{Gopalswamy2009}
	{Gopalswamy}, N., {Yashiro}, S., {Michalek}, G., {et~al.} 2009, Earth Moon and
	Planets, 104, 295, \dodoi{10.1007/s11038-008-9282-7}
	
	\bibitem[{{Gou} {et~al.}(2019){Gou}, {Liu}, {Kliem}, {Wang}, \&
		{Veronig}}]{Gou2019}
	{Gou}, T., {Liu}, R., {Kliem}, B., {Wang}, Y., \& {Veronig}, A.~M. 2019,
	Science Advances, 5, 7004, \dodoi{10.1126/sciadv.aau7004}
	
	\bibitem[{{Gou} {et~al.}(2015){Gou}, {Liu}, \& {Wang}}]{Gou2015}
	{Gou}, T., {Liu}, R., \& {Wang}, Y. 2015, \solphys, 290, 2211,
	\dodoi{10.1007/s11207-015-0750-8}
	
	\bibitem[{{Gou} {et~al.}(2016){Gou}, {Liu}, {Wang}, {Liu}, {Zhuang}, {Chen},
		{Zhang}, \& {Liu}}]{Gou2016}
	{Gou}, T., {Liu}, R., {Wang}, Y., {et~al.} 2016, \apjl, 821, L28,
	\dodoi{10.3847/2041-8205/821/2/L28}
	
	\bibitem[{{Gou} {et~al.}(2017){Gou}, {Veronig}, {Dickson}, {Hernand ez-Perez},
		\& {Liu}}]{Gou2017}
	{Gou}, T., {Veronig}, A.~M., {Dickson}, E.~C., {Hernand ez-Perez}, A., \&
	{Liu}, R. 2017, \apjl, 845, L1, \dodoi{10.3847/2041-8213/aa813d}
	
	\bibitem[{{Hirayama}(1974)}]{Hirayama1974}
	{Hirayama}, T. 1974, \solphys, 34, 323, \dodoi{10.1007/BF00153671}
	
	\bibitem[{{Holman}(2012)}]{Holman2012}
	{Holman}, G.~D. 2012, Physics Today, 65, 56, \dodoi{10.1063/PT.3.1520}
	
	\bibitem[{{Howard} {et~al.}(2008){Howard}, {Moses}, {Vourlidas}, {Newmark},
		{Socker}, {Plunkett}, {Korendyke}, {Cook}, {Hurley}, {Davila}, {Thompson},
		{St Cyr}, {Mentzell}, {Mehalick}, {Lemen}, {Wuelser}, {Duncan}, {Tarbell},
		{Wolfson}, {Moore}, {Harrison}, {Waltham}, {Lang}, {Davis}, {Eyles},
		{Mapson-Menard}, {Simnett}, {Halain}, {Defise}, {Mazy}, {Rochus}, {Mercier},
		{Ravet}, {Delmotte}, {Auchere}, {Delaboudiniere}, {Bothmer}, {Deutsch},
		{Wang}, {Rich}, {Cooper}, {Stephens}, {Maahs}, {Baugh}, {McMullin}, \&
		{Carter}}]{Howard2008}
	{Howard}, R.~A., {Moses}, J.~D., {Vourlidas}, A., {et~al.} 2008, \ssr, 136, 67,
	\dodoi{10.1007/s11214-008-9341-4}
	
	\bibitem[{{Innes} {et~al.}(2014){Innes}, {Guo}, {Bhattacharjee}, {Huang}, \&
		{Schmit}}]{Innes2014}
	{Innes}, D.~E., {Guo}, L.~J., {Bhattacharjee}, A., {Huang}, Y.~M., \& {Schmit},
	D. 2014, \apj, 796, 27, \dodoi{10.1088/0004-637X/796/1/27}
	
	\bibitem[{{Kaiser} {et~al.}(2008){Kaiser}, {Kucera}, {Davila}, {St. Cyr},
		{Guhathakurta}, \& {Christian}}]{Kaiser2008}
	{Kaiser}, M.~L., {Kucera}, T.~A., {Davila}, J.~M., {et~al.} 2008, \ssr, 136, 5,
	\dodoi{10.1007/s11214-007-9277-0}
	
	\bibitem[{{Kopp} \& {Pneuman}(1976)}]{KP1976}
	{Kopp}, R.~A., \& {Pneuman}, G.~W. 1976, \solphys, 50, 85,
	\dodoi{10.1007/BF00206193}
	
	\bibitem[{{Lamy} {et~al.}(2019){Lamy}, {Floyd}, {Boclet}, {Wojak}, {Gilardy},
		\& {Barlyaeva}}]{Lamy2019}
	{Lamy}, P.~L., {Floyd}, O., {Boclet}, B., {et~al.} 2019, \ssr, 215, 39,
	\dodoi{10.1007/s11214-019-0605-y}
	
	\bibitem[{{Lemen} {et~al.}(2012){Lemen}, {Title}, {Akin}, {Boerner}, {Chou},
		{Drake}, {Duncan}, {Edwards}, {Friedlaender}, {Heyman}, {Hurlburt}, {Katz},
		{Kushner}, {Levay}, {Lindgren}, {Mathur}, {McFeaters}, {Mitchell}, {Rehse},
		{Schrijver}, {Springer}, {Stern}, {Tarbell}, {Wuelser}, {Wolfson}, {Yanari},
		{Bookbinder}, {Cheimets}, {Caldwell}, {Deluca}, {Gates}, {Golub}, {Park},
		{Podgorski}, {Bush}, {Scherrer}, {Gummin}, {Smith}, {Auker}, {Jerram},
		{Pool}, {Soufli}, {Windt}, {Beardsley}, {Clapp}, {Lang}, \&
		{Waltham}}]{Lemen2012}
	{Lemen}, J.~R., {Title}, A.~M., {Akin}, D.~J., {et~al.} 2012, \solphys, 275,
	17, \dodoi{10.1007/s11207-011-9776-8}
	
	\bibitem[{{Lin}(2004)}]{LinJ2004}
	{Lin}, J. 2004, \solphys, 222, 115, \dodoi{10.1023/B:SOLA.0000036875.14102.39}
	
	\bibitem[{{Lin} \& {Forbes}(2000)}]{Lin2000}
	{Lin}, J., \& {Forbes}, T.~G. 2000, \jgr, 105, 2375,
	\dodoi{10.1029/1999JA900477}
	
	\bibitem[{{Lin} {et~al.}(2005){Lin}, {Ko}, {Sui}, {Raymond}, {Stenborg},
		{Jiang}, {Zhao}, \& {Mancuso}}]{Lin2005}
	{Lin}, J., {Ko}, Y.~K., {Sui}, L., {et~al.} 2005, \apj, 622, 1251,
	\dodoi{10.1086/428110}
	
	\bibitem[{{Lin} {et~al.}(2002){Lin}, {Dennis}, {Hurford}, {Smith}, {Zehnder},
		{Harvey}, {Curtis}, {Pankow}, {Turin}, {Bester}, {Csillaghy}, {Lewis},
		{Madden}, {van Beek}, {Appleby}, {Raudorf}, {McTiernan}, {Ramaty}, {Schmahl},
		{Schwartz}, {Krucker}, {Abiad}, {Quinn}, {Berg}, {Hashii}, {Sterling},
		{Jackson}, {Pratt}, {Campbell}, {Malone}, {Landis}, {Barrington-Leigh},
		{Slassi-Sennou}, {Cork}, {Clark}, {Amato}, {Orwig}, {Boyle}, {Banks},
		{Shirey}, {Tolbert}, {Zarro}, {Snow}, {Thomsen}, {Henneck}, {McHedlishvili},
		{Ming}, {Fivian}, {Jordan}, {Wanner}, {Crubb}, {Preble}, {Matranga}, {Benz},
		{Hudson}, {Canfield}, {Holman}, {Crannell}, {Kosugi}, {Emslie}, {Vilmer},
		{Brown}, {Johns-Krull}, {Aschwanden}, {Metcalf}, \& {Conway}}]{LinRP2002}
	{Lin}, R.~P., {Dennis}, B.~R., {Hurford}, G.~J., {et~al.} 2002, \solphys, 210,
	3, \dodoi{10.1023/A:1022428818870}
	
	\bibitem[{{Liu}(2013)}]{LiuR2013}
	{Liu}, R. 2013, \mnras, 434, 1309, \dodoi{10.1093/mnras/stt1090}
	
	\bibitem[{{Liu} {et~al.}(2013){Liu}, {Chen}, \& {Petrosian}}]{LiuW2013}
	{Liu}, W., {Chen}, Q., \& {Petrosian}, V. 2013, \apj, 767, 168,
	\dodoi{10.1088/0004-637X/767/2/168}
	
	\bibitem[{{Mann} {et~al.}(2003){Mann}, {Klassen}, {Aurass}, \&
		{Classen}}]{Mann2003}
	{Mann}, G., {Klassen}, A., {Aurass}, H., \& {Classen}, H.~T. 2003, \aap, 400,
	329, \dodoi{10.1051/0004-6361:20021593}
	
	\bibitem[{{Mari{\v{c}}i{\'c}} {et~al.}(2007){Mari{\v{c}}i{\'c}},
		{Vr{\v{s}}nak}, {Stanger}, {Veronig}, {Temmer}, \&
		{Ro{\v{s}}a}}]{Maricic2007}
	{Mari{\v{c}}i{\'c}}, D., {Vr{\v{s}}nak}, B., {Stanger}, A.~L., {et~al.} 2007,
	\solphys, 241, 99, \dodoi{10.1007/s11207-007-0291-x}
	
	\bibitem[{{Mart{\'\i}nez Oliveros} {et~al.}(2014){Mart{\'\i}nez Oliveros},
		{Krucker}, {Hudson}, {Saint-Hilaire}, {Bain}, {Lindsey}, {Bogart},
		{Couvidat}, {Scherrer}, \& {Schou}}]{MO2014}
	{Mart{\'\i}nez Oliveros}, J.-C., {Krucker}, S., {Hudson}, H.~S., {et~al.} 2014,
	\apjl, 780, L28, \dodoi{10.1088/2041-8205/780/2/L28}
	
	\bibitem[{{McKenzie} \& {Hudson}(1999)}]{McKenzie1999}
	{McKenzie}, D.~E., \& {Hudson}, H.~S. 1999, \apjl, 519, L93,
	\dodoi{10.1086/312110}
	
	\bibitem[{{Pesnell} {et~al.}(2012){Pesnell}, {Thompson}, \&
		{Chamberlin}}]{Pesnell2012}
	{Pesnell}, W.~D., {Thompson}, B.~J., \& {Chamberlin}, P.~C. 2012, \solphys,
	275, 3, \dodoi{10.1007/s11207-011-9841-3}
	
	\bibitem[{{Podladchikova} {et~al.}(2017){Podladchikova}, {Van der Linden}, \&
		{Veronig}}]{Podl2017}
	{Podladchikova}, T., {Van der Linden}, R., \& {Veronig}, A.~M. 2017, \apj, 850,
	81, \dodoi{10.3847/1538-4357/aa93ef}
	
	\bibitem[{{Priest} \& {Forbes}(2002)}]{Priest2002}
	{Priest}, E.~R., \& {Forbes}, T.~G. 2002, \aapr, 10, 313,
	\dodoi{10.1007/s001590100013}
	
	\bibitem[{{Saint-Hilaire} {et~al.}(2014){Saint-Hilaire}, {Schou},
		{Mart{\'\i}nez Oliveros}, {Hudson}, {Krucker}, {Bain}, \&
		{Couvidat}}]{SH2014}
	{Saint-Hilaire}, P., {Schou}, J., {Mart{\'\i}nez Oliveros}, J.-C., {et~al.}
	2014, \apjl, 786, L19, \dodoi{10.1088/2041-8205/786/2/L19}
	
	\bibitem[{{Savage} \& {McKenzie}(2011)}]{Savage2011}
	{Savage}, S.~L., \& {McKenzie}, D.~E. 2011, \apj, 730, 98,
	\dodoi{10.1088/0004-637X/730/2/98}
	
	\bibitem[{{Shibata} \& {Magara}(2011)}]{Shibata2011}
	{Shibata}, K., \& {Magara}, T. 2011, Living Reviews in Solar Physics, 8, 6,
	\dodoi{10.12942/lrsp-2011-6}
	
	\bibitem[{{Sturrock}(1966)}]{Sturrock1966}
	{Sturrock}, P.~A. 1966, \nat, 211, 695, \dodoi{10.1038/211695a0}
	
	\bibitem[{{Temmer} {et~al.}(2010){Temmer}, {Veronig}, {Kontar}, {Krucker}, \&
		{Vr{\v{s}}nak}}]{Temmer2010}
	{Temmer}, M., {Veronig}, A.~M., {Kontar}, E.~P., {Krucker}, S., \&
	{Vr{\v{s}}nak}, B. 2010, \apj, 712, 1410,
	\dodoi{10.1088/0004-637X/712/2/1410}
	
	\bibitem[{{Temmer} {et~al.}(2008){Temmer}, {Veronig}, {Vr{\v{s}}nak},
		{Ryb{\'a}k}, {G{\"o}m{\"o}ry}, {Stoiser}, \&
		{Mari{\v{c}}i{\'c}}}]{Temmer2008}
	{Temmer}, M., {Veronig}, A.~M., {Vr{\v{s}}nak}, B., {et~al.} 2008, \apjl, 673,
	L95, \dodoi{10.1086/527414}
	
	\bibitem[{{Thernisien} {et~al.}(2009){Thernisien}, {Vourlidas}, \&
		{Howard}}]{Thernisien2009}
	{Thernisien}, A., {Vourlidas}, A., \& {Howard}, R.~A. 2009, \solphys, 256, 111,
	\dodoi{10.1007/s11207-009-9346-5}
	
	\bibitem[{{Thernisien} {et~al.}(2006){Thernisien}, {Howard}, \&
		{Vourlidas}}]{Thernisien2006}
	{Thernisien}, A.~F.~R., {Howard}, R.~A., \& {Vourlidas}, A. 2006, \apj, 652,
	763, \dodoi{10.1086/508254}
	
	\bibitem[{{Tsuneta}(1996)}]{Tsuneta1996}
	{Tsuneta}, S. 1996, \apj, 456, 840, \dodoi{10.1086/176701}
	
	\bibitem[{{Veronig} {et~al.}(2018){Veronig}, {Podladchikova}, {Dissauer},
		{Temmer}, {Seaton}, {Long}, {Guo}, {Vr{\v{s}}nak}, {Harra}, \&
		{Kliem}}]{Veronig2018}
	{Veronig}, A.~M., {Podladchikova}, T., {Dissauer}, K., {et~al.} 2018, \apj,
	868, 107, \dodoi{10.3847/1538-4357/aaeac5}
	
	\bibitem[{{Vr{\v{s}}nak}(2008)}]{Vrsnak2008}
	{Vr{\v{s}}nak}, B. 2008, Annales Geophysicae, 26, 3089,
	\dodoi{10.5194/angeo-26-3089-2008}
	
	\bibitem[{{Vr{\v{s}}nak} {et~al.}(2004){Vr{\v{s}}nak}, {Mari{\v{c}}i{\'c}},
		{Stanger}, \& {Veronig}}]{Vrsnak2004}
	{Vr{\v{s}}nak}, B., {Mari{\v{c}}i{\'c}}, D., {Stanger}, A.~L., \& {Veronig}, A.
	2004, \solphys, 225, 355, \dodoi{10.1007/s11207-004-4995-x}
	
	\bibitem[{{Vr{\v{s}}nak} {et~al.}(2007){Vr{\v{s}}nak}, {Mari{\v{c}}i{\'c}},
		{Stanger}, {Veronig}, {Temmer}, \& {Ro{\v{s}}a}}]{Vrsnak2007}
	{Vr{\v{s}}nak}, B., {Mari{\v{c}}i{\'c}}, D., {Stanger}, A.~L., {et~al.} 2007,
	\solphys, 241, 85, \dodoi{10.1007/s11207-006-0290-3}
	
	\bibitem[{{Vr{\v{s}}nak} {et~al.}(2006){Vr{\v{s}}nak}, {Temmer}, {Veronig},
		{Karlick{\'y}}, \& {Lin}}]{Vrsnak2006}
	{Vr{\v{s}}nak}, B., {Temmer}, M., {Veronig}, A., {Karlick{\'y}}, M., \& {Lin},
	J. 2006, \solphys, 234, 273, \dodoi{10.1007/s11207-006-0093-6}
	
	\bibitem[{{Warren} {et~al.}(2018){Warren}, {Brooks}, {Ugarte-Urra}, {Reep},
		{Crump}, \& {Doschek}}]{Warren2018}
	{Warren}, H.~P., {Brooks}, D.~H., {Ugarte-Urra}, I., {et~al.} 2018, \apj, 854,
	122, \dodoi{10.3847/1538-4357/aaa9b8}
	
	\bibitem[{{Zhang} \& {Dere}(2006)}]{ZhangJ2006}
	{Zhang}, J., \& {Dere}, K.~P. 2006, \apj, 649, 1100, \dodoi{10.1086/506903}
	
	\bibitem[{{Zhang} {et~al.}(2001){Zhang}, {Dere}, {Howard}, {Kundu}, \&
		{White}}]{ZhangJ2001}
	{Zhang}, J., {Dere}, K.~P., {Howard}, R.~A., {Kundu}, M.~R., \& {White}, S.~M.
	2001, \apj, 559, 452, \dodoi{10.1086/322405}
	
	\bibitem[{{Zhang} {et~al.}(2004){Zhang}, {Dere}, {Howard}, \&
		{Vourlidas}}]{ZhangJ2004}
	{Zhang}, J., {Dere}, K.~P., {Howard}, R.~A., \& {Vourlidas}, A. 2004, \apj,
	604, 420, \dodoi{10.1086/381725}
	
\end{thebibliography}
\end{document}